# Computational Security Analysis of the UMTS and LTE Authentication and Key Agreement Protocols

Joe-Kai Tsay and Stig F. Mjølsnes

Department of Telematics
Norwegian University of Sciences and Technology, NTNU
{joe.k.tsay,sfm@item.ntnu.no}

**Abstract.** One of the forerunners and main candidates for the fourth generation (4G) generation mobile communication system is commonly known under the name Long-Term Evolution (LTE) and its standard is produced and maintained by the international 3rd Generation Partnership Program (3GPP) consortium. The LTE Authentication and Key Agreement (AKA) protocol design is based on the Universal Mobile Telecommunications System (UMTS) AKA protocol, which is widely used today for third generation (3G) wireless networks, and which itself is the successor of the Subscriber Identity Authentication (SIA) protocol of the Global System for Mobile Communication (GSM). With the persistent spread of these mobile network systems, their authentication protocols have become some of the most widely used security protocols today. We present a computational security analysis of both the LTE AKA and the UMTS AKA. This work constitutes the first security analysis of LTE AKA to date and the first computationally sound analysis of UMTS AKA. Our work is the first formal analysis to consider messages that are sent in the core network, where we take into account details of the carrying protocol (i.e., MAP or Diameter) and of the mechanism for secure transport (i.e., MAPsec/TCAPsec or IPsec ESP). Moreover, we report on a deficiency in the protocol specifications of UMTS AKA and LTE AKA and the specifications of the core network security (called *network domain security*), which may enable efficient attacks. The vulnerability can be exploited by both an outside and an inside attacker who can violate entity authentication properties. It allows an inside attacker not only to impersonate an honest protocol participant during a run of the protocol but also to subsequently use wireless services on his behalf. UMTS AKA run over MAP with MAPsec seems vulnerable in the most straight-forward application of the attack. On the other hand, our analysis shows that UMTS and LTE AKA over Diameter/IPsec and UMTS AKA over MAP/TCAPsec (with sufficiently long session identifiers) computationally satisfy intended authentication properties as well as some key secrecy properties, assuming that the used primitives meet standard cryptographic assumptions.

## 1 Introduction

These are exciting times in the development of mobile networks. The GSM and UMTS mobile networks are a worldwide success with now about 6 billion supscriptions [40], and still growing. The pace of innovation in mobile terminal hardware and software is amazing, where touch screen and online packet switched internet access have become a consumer standard. New mobile systems are rolled out, which include the 3GPP recent developments named 'Long Term Evolution' (LTE) and 'System Architecture Evolution' (SAE). The new system is called 'Evolved Packet System (EPS)', emphasizing the all-IP packet switching design throughout the system unto the user's mobile terminal. Although EPS is the proper technical term for this new 3GPP mobile system generation of SAE/LTE, we will keep with the most well-known name LTE. There is a multitude of security issues in such large networked systems. Here we will focus on the mobile terminal access security by means of an authentication and key agreement protocol (AKA) that is structured very similar in UMTS and LTE. The technical problem we are addressing is to find out whether the UMTS AKA and LTE AKA are secure with respect to the *Computational Model*, *i.e.* the security model that is used in *Modern Cryptography* [35,36]. This work constitutes the first computational analysis of both UMTS AKA and LTE AKA. While there exist formal analyses of UMTS AKA in the *Symbolic Model* of security (also called the *Dolev-Yao* model and inspired by [34]), it is in fact the first security analysis of LTE AKA to date. While there exist already formal security proofs for UMTS AKA in the symbolic model, this work provides the first security proofs for LTE AKA to date. Although the design of LTE AKA is based on UMTS AKA, its security can a priori not be deduced from the security properties of UMTS AKA as, for instance, the LTE AKA protocol has been designed to offer stronger authentication guarantees. Furthermore, our proofs are the first that cover the AKA[1] messages sent within the core networks, where we take into account security details of the protocols over which the UMTS and LTE AKA protocol messages are transported. Namely, our proofs examine security properties of the UMTS and LTE AKA protocol for the scenario with the strongest protection described in the UMTS and LTE specifications: within the core network, the AKA messages are carried over the Mobile Application Part (MAP) protocol [2] or the Diameter protocol [3,37] and are additionally protected via MAPsec [5] or TCAPsec [6] (which is the descendant of MAPsec) or IPsec ESP [7]. We note that there are scenarios where the 3GPP specifications [7,5,6] allow network operators to use their proprietary solutions for protecting core network messages or where the AKA protocols can be executed without any confidentiality protection. Proprietary solutions can generally not be verified unless explicit details are given[2]. And in the case that no confidentiality protection is used in the core network, the exchanged session keys are exposed and can be eavesdropped, and so there are no meaningful security properties to prove.

Designing security protocols has proven to be a very error-prone task. There exist many examples of security protocols that were believed to be secure, usually with respect to the symbolic or the computational model of security, but that were later shown to be flawed (often in its logical structure) [47,48,52,29]. Especially ensuring the security of a protocol under concurrent executions of multiple protocol sessions (*runs*) is a challenging task in analyzing security protocols by hand. Therefore, considerable effort has been directed towards the development of tools that can analyze security protocols in an automatic fashion. There are already several tools that are able to analyze security protocols with respect to the symbolic model. In particular secrecy and authentication properties, even of complex protocols, can be shown with a high degree of automation. But also other properties, *e.g.* privacy and verifiability properties of electronic voting protocols, can be handled by such symbolic provers [32,15,18,22,49]. While attacks found against protocols within the symbolic model, where cryptographic primitives are perfectly secure, immediately translate to attacks in the computational model, it is not clear what the successful verification of a protocol in the symbolic model signifies when the security of the protocol is considered in the more fine-grained computational model where an adversary may attack a protocol via the used primitives. In order to take advantage of the existing effective tools that work in the symbolic model but, at the same time, obtain the stronger security guarantees of the computational model, a line of research focuses on so-called *Computationally Sound* frameworks, *e.g.* [16,28,33,31,30]. However, currently such frameworks lack the tool support yet which would allow proofs without a lot of user interaction. In recent years, an alternative direction has been taken by building tools that prove the security of protocols or cryptographic primitives directly in the computational model [25,17]. One such solution, CryptoVerif, has previously been applied to obtain mechanized security proofs in the computational model of complex real-world protocols such as Kerberos [26] or TLS [21].

We also conduct our analysis of UMTS AKA and LTE AKA with CryptoVerif. Our results are twofold: On the one hand, we discover a previously undetected vulnerability in the specifications of both UMTS AKA and LTE AKA and the specifications of the core network security [7,5,6].

---

[1] We sometimes write only *AKA* when we talk about both UMTS AKA and LTE AKA
[2] We do not have any knowledge about proprietary solutions (and their specifics) for protecting core network AKA messages that deviate from the strongest protection described in the specifications. (and considered in this work)

Due to an identity misbinding of the authentication vectors an attacker can execute a parallel session attack.[3] The vulnerability found could be exploited by both outside and inside attackers in order to break authentication of a user to a serving network. Furthermore, inside attackers may impersonate an honest user and use wireless services on his behalf without the user being present on the network at that time. We reported the vulnerability to the 3GPP in Spring 2012 and have since been waiting for any reply. In practice, UMTS AKA over MAP/MAPsec is flawed as the session identifiers (called *TCAP transaction identifier*) are not protected. And the attack may also be effective against both UMTS AKA and LTE AKA over intra-domain, depending on the proprietary security solutions in place.

On the other hand, our analysis shows that UMTS and LTE AKA run over Diameter/IPsec EPS and UMTS over MAP/TCAPsec are computationally secure: Under standard assumptions on the cryptographic primitives, *e.g.*, including IND-CPA secure symmetric encryption and WUF-CMA secure integrity protection, and under the additional assumption that the session identifier used in the carrying protocols (MAP or Diameter) are unique with overwhelming probability, we use CryptoVerif to prove intended authentication properties and secrecy properties for the session keys that are exchanged with the AKAs.

**Related Work** Annex B of the 3GPP technical report TR33.902 (2001) [1] documents a formal analysis of the UMTS AKA protocol using a BAN logic variant, *i.e.*, a symbolic and not computational analysis. The analysis verifies authentication and secrecy properties under the assumption that the home network acts as a trusted third party. The flaw that we present here is not detected in [1] because strong assumptions (called *prerequisites on SN's side*) are used which already eliminate the weakness in the protocol and the network domain security.

The GSM Subscriber Identity Authentication protocol does not provide for the authentication of the access network, which obviously creates a problem with detecting false base stations. The interoperability of the GSM and UMTS systems perpetuates this attack possibility, reported in [50]. Our analysis is not directed to the problems of interoperability between LTE/UMTS/GSM.

[54] formally analyze the security UMTS and GSM roaming protocols using the tool ProVerif [22] with respect to the symbolic security model. Again, they use strong assumptions on the core network communication between serving networks and home networks (namely, they use private channels), which cause them to overlook the vulnerability we discover.

A redirection attack on the UMTS AKA is reported in [55], which exploits the observation that the user is not able to authenticate the identity of the *serving* network because this is not included in the authentication vector provided by the home network. The new LTE AKA specification that we analyze is designed to fix this weakness and implicitly authenticate the serving network to the user.

A recent paper focuses on the unlinkability and anonymity properties of the protocol [14]. They also use the ProVerif tool for a symbolic analysis, and the paper describes an attack that enables the adversary to distinguish a known user from any other. This is done by replaying captured messages from the known user and use the different error messages that are returned. The analysis models the UMTS AKA as a simplified two-party protocol between a user and the core network. However, by reducing UMTS AKA to a two-party protocol, the weakness uncovered in the present work is stamped out.

Our work is also relevant for RFC 4187 [39] that specifies an *Extensible Authentication Protocol Method for 3rd Generation Authentication and Key Agreement (EAP-AKA)* based on UMTS AKA, which is compatible with [38]. In particular, close attention needs to be paid in case the communication between EAP authenticator and authentication server should be secured. We note that the authors of [39] explicitly state that the IETF has not validated the security claims.

**Structure of this work** In Section 2, we will give an overview of the Mobile Network architecture and give a description of the UMTS AKA and LTE AKA protocols. In Section 3, we describe the vulnerability we found on both on the specifications of UMTS and LTE AKA and its consequences. In Section 4, we present our computational security proofs of UMTS and LTE AKA using the tool CryptoVerif. Finally, we conclude with Section 5.

---

[3] We note that the attack is also relevant for the cases where the protocol messages of the GSM Subscriber Identity Authentication [12,11] are meant to be protected within the core network (cf. Appendix B).

## 2 The UMTS and LTE Authentication and Key Agreement Protocols

### 2.1 Overview of the Mobile Network Architecture

For both UMTS and LTE the basic network architectures are very similar. In comparison to UMTS, the network elements used for LTE are upgraded and mostly renamed. However, they fulfill the analogous tasks in both cases. In order to avoid unnecessary confusion over terminology, we give a unified description of the network architectures of UMTS and LTE at the level of detail necessary for understanding our analysis presented below. Basically, the mobile network architecture comprises three parts, that is, the user's mobile equipment $U$, the Radio Access Network (RAN), and the Core Network (CN). The user equipment consists of the mobile equipment and a tamper-resistant chip card, the *Universal Subscriber Identity Module* (USIM). The USIM is issued by a mobile operator to a subscriber and contains the International Mobile Subscriber Identity (IMSI), the permanent key of the subscription shared between subscriber and operator, and the cryptographic algorithms for the authentication protocol. In the following, we will use the terms *user*, *subscriber* and *user equipment* interchangeably. Each mobile operator runs an *Authentication Center* (*AuC*) server within its core network that contains the security related information of all the subscribers of the operator and generates temporary security credentials to be used by a user and a core network to establish authentication guarantees and set up session keys. The core network is divided into a *serving network S* and a *home network H*, where the latter contains and maintains the AuC and the serving network is responsible for the communication to the user equipment through the radio access network.

The serving network and the home network do not necessarily belong to the same security domain, *i.e.*, they may be controlled by different mobile operators. A subscriber $U_1$ of a mobile operator $OP_1$ with home network $H_1$ may *roam* into the domain of mobile operator $OP_2$'s radio access network maintained by serving network $S_2$. If $OP_1$ has a roaming agreement with $OP_2$, then $U_1$ will be able to access the mobile network through $S_2$'s radio access network. In this case, the connections between $S_2$ and $H_1$ are called *inter-domain connections*. In comparison, the connections within a core network controlled by a single mobile operator, *i.e.*, between $S_i$ and $H_i$, for $i \in \{1, 2\}$, are called *intra-domain connections*.

### 2.2 UMTS & LTE AKA

Figure 1 shows the message sequence diagram description of the authentication and key agreement protocol in a unified way for UMTS and LTE on a similar level of detail as depicted in [4,9]. The protocol is executed between user $U$, visited serving network $S$ and $U$'s home network $H$. $U$ and $H$ share the long-term key $K$ and a set of algorithms $f_1, \ldots, f_4$ and, in the case of LTE, also a key derivation function $KDF$. The functions $f_1, f_2$ are so called *message authentication functions*, and $f_3, f_4$ are so called *key generating functions*[4]. Moreover, $U$ maintains a counter $SQN_U$ and $H$ a counter $SQN_H$ for $U$.

A protocol run starts with $S$ sending a *user id request* and $U$ responding with its *IMSI*[5]. Next follows the *authentication data transfer*, in which $S$ sends an *authentication data request* to $H$, that consists of $U$'s *IMSI* and $S$'s identifier *SNid*, and $H$ answers with an *authentication data response*. $H$ chooses a fresh nonce *RAND* and computes, with the key $K$ and its sequence number $SQN_H$, the so-called *message authentication code MAC*, the *expected response XRES*, the *cipher* key $CK$, the *integrity key* $IK$, and the *authentication token* $AUTN$ as depicted in Figure 1, where $\|$ denotes concatenation. The main difference between the UMTS AKA and LTE AKA is the *session key Skey*. In LTE AKA, the session key is computed over the identifier of $S$. There is the option that $H$ sends $S$ multiple *authentication vectors* $(RAND_i, AUTN_i, XRES_i, Skey_i)$ for $i = 1, \ldots, n$ at once in order to reduce the traffic between $S$ and $H$.

---
[4] We choose to do without the *anonymity key*, i.e. $f_5 \equiv 0$, which is an option in the specifications. We also omit the AMF constant.
[5] In fact, $U$ may alternatively respond with a temporary mobile subscriber identity (TMSI), which reduces but does not fully avoid the use of the *IMSI*.

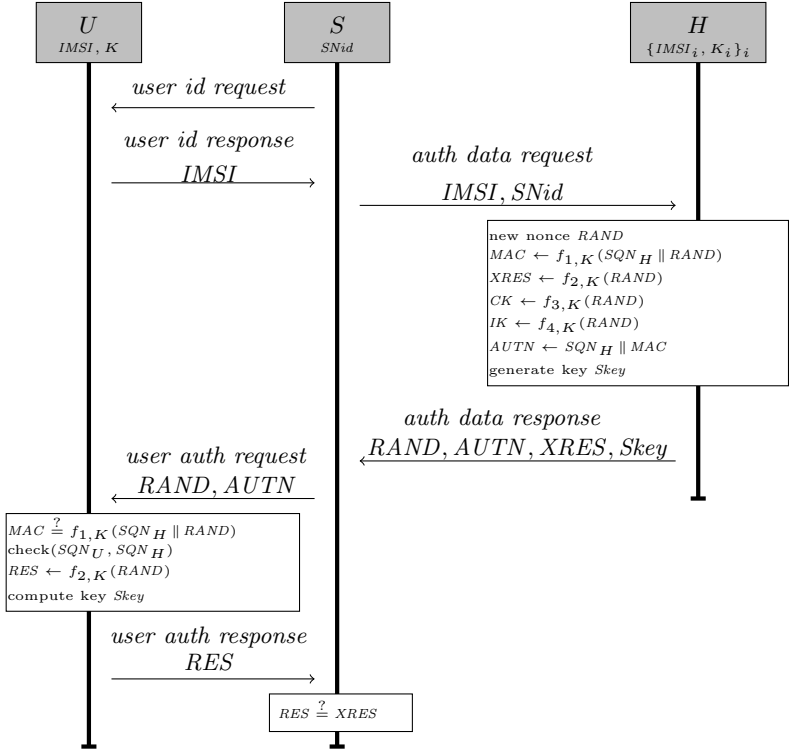

**Fig. 1.** The UMTS/LTE Authentication and Key Agreement Protocol. The session key in UMTS is $Skey \leftarrow CK \,\|\, IK$, and in LTE it is $Skey := K_{ASME} \leftarrow KDF(SQN_H \,\|\, CK \,\|\, IK \,\|\, SNid)$.

In the *user authentication request*, $S$ forwards only $RAND$ and $AUTN$ to $U$. From the received $RAND, AUTN$, the user $U$ extracts $SQN_H$, computes the *expected message authentication code XMAC* and compares it to $MAC$ contained in $AUTN$. If they are equal then $U$ performs a check on the sequence numbers $SQN_H$ and $SQN_U$[6]. If either of this two checks fail, then $U$ sends some error messages to $S$ (in fact, the error messages may be different, therefore allowing the linkability attack of [14]). Otherwise $U$ computes the *response RES* and sends it to $S$. User $U$ can compute the session key *Skey* from $RAND$ and $K$. Finally, $S$ compares the response received from $U$ with the expected response received from $H$; if they are equal then the UMTS/LTE AKA run was successfully completed.

Intuitively, the UMTS/LTE AKA establishes the session key *Skey* between $U$ and $S$, therefore, *Skey* must satisfy some secrecy property. Furthermore, the protocol aims to authenticate $U$ to $S$. Both properties require $S$ to trust $H$ to provide a correct authentication data response. The sequence numbers allow to detect possible replays of authentication tokens. The UMTS/LTE AKA protocol, as depicted in Figure 1, does not offer authentication of $S$ to $U$. This known weakness has been described in [55]. User $U$ may at most know that $H$ generated the received nonce and authentication token for some service network.

Following the UMTS/LTE AKA, serving network $S$ and user $U$ need to negotiate the cryptographic algorithms (*security mode*) used to protect subsequent wireless communication between $S$ and $U$. Note that these algorithms are, in particular for inter-domain connections, not predetermined. The messages of this negotiation are protected by (keys derived from) *Skey*. This is especially relevant for the case of LTE, where *Skey* is generated over $S$'s identifier *SNid*. In LTE, by receiving the *NAS security mode command* directly following the user authentication response of the AKA, $U$ should be able to authenticate $S$, as this message constitutes a key confirmation of the session key $K_{ASME}$. According to [9], the NAS security mode command sent from $S$ to $U$ has following form:

$$S \longrightarrow U : eKSI, UE\ security\ capabil., ciph.\ algo, int.\ algo, NAS\text{-}MAC$$

where *NAS-MAC* is a message authentication code under a key derived from $K_{ASME}$ over the rest of the message, which consists of non-secret components[7]. We denote by *LTE AKA+1* the LTE AKA protocol together with this NAS security mode command message.

---

[6] Checking and increasing the sequence numbers can be done in different ways

## 3 A Vulnerability in UMTS & LTE AKA

Here we present a weakness found in the authentication protocol specifications of both UMTS and LTE AKA with the help of the tool CryptoVerif [23]. Although CryptoVerif has semantics in the computational model, the flaw in the protocols is of symbolic nature. Unlike other provers that work in the symbolic model, CryptoVerif does not output attack traces; instead we found the attack by interpreting the *last game* in a sequence of game transformations performed by CryptoVerif. It is the same flaw that is present in the specifications of both UMTS AKA and LTE AKA. Although UMTS AKA has previously been formally analyzed [14,1], none of the previous analyses have detected this flaw. How GSM SIA is affected by the flaw is discussed in Appendix B.

### 3.1 Communication Security Between $S$ and $H$

As given in Figure 1, the communication in the core network, *i.e.*, between $S$ and $H$, is not protected even though it can involve long-distance signalling, *e.g.*, over IP networks. However it is obvious that the communication between $S$ and $H$ should be protected in some way against a network attacker, otherwise the exchanged session key(s) are sent in the clear. The specifications of the security architectures of UMTS and LTE in [9] and [4], which specify the AKA protocols, mention little about the security protection of the authentication data transfer. For instance, the AKA protocols are given in [9] and [4] at essentially the same level of detail as in Figure 1. However, for UMTS and LTE, the *Network Domain Security* (NDS) specifications detail the protection of core network communication: [7,8] specify the protection of IP-based communication between network elements. In addition, for UMTS, the communication between $S$ and $H$ can also be carried out on the *global SS7 network* and its protection is then specified in [6,5]. In all cases, the specifications distinguish between inter-domain communication, where communicating parties are not controlled by the same mobile operator, and intra-domain communication, where the communicating parties are controlled by the same mobile operator.

For inter-domain connections over IP-based networks, [7,8] mandate the protection of the communication between network elements using IPsec with Encapsulating Security Payload (ESP) mode. For inter-domain connections over SS7 networks, the specification [6] mandates the protection using *Transaction Capabilities Application Part* security (TCAPsec), which is the successor for *Mobile Application Part security* (MAPsec) [5]. According to the specifications [5,6,7], IPsec ESP, MAPsec and TCAPsec should all provide

- *data integrity*
- *data origin authentication*
- *anti-replay protection*

---

[7] *eKSI* is the key identifier for $K_{ASME}$, *UE security capabil* are the security capabilities that *UE* had sent to $S$ before running the AKA, *ciph. algo* are the algorithms chosen by the serving network for encryption, *int. algo* are the algorithms chosen by the serving network for integrity protection

– *confidentiality (optional)*

For the case of IPsec ESP, [7] also lists as security guarantee

– *limited protection against traffic flow analysis when confidentiality is applied.*

For intra-domain connections over IP-based networks or SS7 networks, however, [7,8,6,5] state that the protection of communication is regarded as an internal issue of each domain operator. In particular, utilizing MAPsec, TCAPsec or IPsec ESP for intra-domain communication between $S$ and $H$ is *optional*, even though the communication may involve long distance signaling. Instead it is the decision of the operator what kind of protection of the core network communication he/she wants in the network domain and which (proprietary) solutions he/she may deploy to achieve that protection.

### 3.2 Session-mixup Attack against the Authentication Data Response

For our attacks on the UMTS and LTE AKA we consider, as usual, an adversary who is in full control of the messages sent between instances of the roles of user $U$, serving network $S$, and home network $H$. In particular, the adversary can control all incoming and outgoing messages of a serving network $S$. We assume that the home network $H$ acts as a trusted third party. We assume that the messages sent between $S$ and $H$ are encrypted and then integrity protected through a message authentication code under long-term keys shared between $S$ and $H$(notice that support for pre-shared keys is required in [5,6,7]), while not making any distinction between intra- or inter-domain connections. The *encrypt-then-mac* scheme is indeed the principle used by IPsec, MAPsec, and TACAPsec. Neglecting momentarily the details and differences of the carrying protocols, we consider the scenario in which two user equipments $U$ and $U'$ are running concurrent sessions with the same serving network $S$. Notice that when $S$ sends an authentication data request to $H$ for authentication parameters of $U$, the authentication data response by $H$ to $S$ is bound to $U$ as it includes message components that are generated under the long-term key shared between $H$ and $U$. However, $S$ cannot verify for which user equipment the received randomness and session key were generated for, as $S$ does not know the key shared between the user equipments and $H$. We present two alternative scenarios in which an attacker may take advantage of this.

**An *Inside* Attack** In this scenario we consider an attacker $A$ who is a subscriber $U$ of $H$. The message flow of this attack is depicted in Figure 2, where we omit the user identity request by $S$ (cf. Figure 1). Say $U'$ is another subscriber of $H$ who is honest. If $A$ knows the $IMSI'$ of $U'$, which $A$ can learn either by listening on the network or by deploying a device called *imsi catcher*, then $A$ can execute the attack that is depicted in Figure 2 without $U'$ even being present. In this case, $A$ does not need to be able to intercept messages sent over the base stations. The attacker sends out two user identity responses: $IMSI'$ and his own subscriber identity $IMSI$. Then $S$ will run two concurrent AKA sessions, one for $U$ and one for $U'$, and sends two authentication data requests to $H$. When $H$ sends the authentication data responses for $S$ and $U$, then adversary $A$ redirects this message such that it is mistaken by $S$ as the response by $H$ for $S$ and $U'$ while he blocks the authentication data response that $H$ generated for $S$ and $U'$. Notice that this *session mixup* can be created by the attacker without breaking any cryptographic primitive and does generally not violate the specifications. Next the attacker redirects the messages sent by $S$ intended for $U'$ to $U$. So $U$ correctly receives the user authentication request containing message components that were generated by $H$ for $U$ (and $S$). Therefore, attacker $A$, who is registered as $U$, can generate the correct response and relay it to to $S$ such that $S$ believes that the response was generated by $U'$. The other session that $S$ opened for $U'$ is halted by $A$; it cannot be completed because $A$ does not know the keys that $U'$ shares with $H$. Anyhow, $A$ can impersonate $U'$ to $S$, therefore, breaking entity authentication. Furthermore, the attacker and $S$ share a session key; it was in fact generated by $H$ for $U$ and $S$. At the same time, $S$ believes that this session key was generated by $H$ for $S$ and $U'$. Therefore, the attacker is able to execute subsequent communication steps and

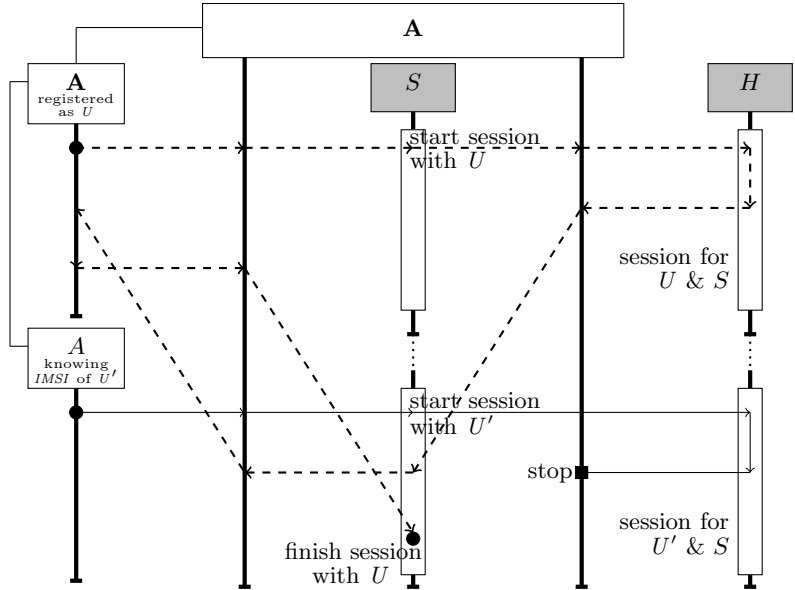

**Fig. 2.** Message flow of an *inside* attack against UMTS and LTE AKA (not showing the user id request). The attacker impersonates honest user $U'$ to $S$ and shares the session key(s) with $S$, without $U'$ being involved.

use the derived keys to use the wireless service provided by $S$ on behalf of $U'$.[8] $S$ will bill $H$ for the service that attacker $A$ received on $U'$ behalf, and $H$ will bill $U'$. Furthermore, an attacker can falsely cause $U'$ to appear physically present within a certain network cell

**An *Outside* Attack** In this attack scenario the attacker $A$ does not need to be a subscriber himself. The message flow of this attack is depicted in Figure 3, where we omit the user identity request by $S$ (cf. Figure 1). Assume that $U$ and $U'$ are subscribers of $H$ and both execute an AKA run with $S$ at the same time. First adversary $A$ correctly forwards the user identity response message by $U$, respectively by $U'$, to $S$ and also forwards the authentication data requests by $S$ to $H$. At this point $S$ has started two AKA sessions, one for $U$ and one for $U'$. When $H$ sends the authentication data responses for $U$ and $U'$, adversary $A$ swaps these messages such that the authentication data response for $S$ and $U$ is mistaken by $S$ as the response by $H$ for $S$ and $U'$, i.e. it is accepted by $S$'s session for $U'$. And $A$ does the analogous with the response by $H$ for $S$ and $U'$. Again, this session mixup can be created by the attacker without breaking any cryptographic primitive and does generally not violate the specifications. Next the attacker redirects the messages sent by $S$ intended for $U'$ to $U$ and vice versa. So $U$ correctly receives the user authentication request containing message components that were generated by $H$ for $U$ and $S$ (and the analogous holds for $U'$). Therefore, under the assumption that the tests on the sequence numbers pass, $U$ and $U'$ do not notice the attack and send out the correct user authentication responses to $S$. Finally, $A$ relays the response by $U$ to $S$ such that $S$ believes that the message came from $U'$, and does analogous to the response by $U'$.

Entity authentication of user equipment to $S$ is clearly violated. At the end of the run, $S$ believes that it successfully completed a run with $U$ while it was in fact $U'$ that participated in that run with $S$. Likewise, while $S$ believes that it completed a run of the protocol with $U'$, it was in fact $U$ who participated in that run with $S$. Furthermore, while $S$ believes that it is sharing a session key *Skey* with $U$, it is in fact sharing the session key *Skey* with $U'$. Analogously, while $S$ believes that it is sharing a session key $Skey'$ with $U'$, it is in fact sharing the session key $Skey'$ with $U$. Therefore, the attacker $A$ may continue swapping the subsequent communication between $U$ and $S$ and the one between $U'$ and $S$, which are both protected by (keys derived from) the exchanged session key. In the case that $U$ and $U'$ are not corrupted, the attacker cannot learn the protected data that is transmitted (assuming the cryptographic primitives are sufficiently secure). However, neither $S$ or $U$ or $U'$ notices that $U$ is authenticated to $S$ as $U'$ and $U'$ as $U$.

---

[8] The attack is not fended off by the use of TMSIs. And the attacker's job is simplified in practice if multiple authentication vectors are sent at once.

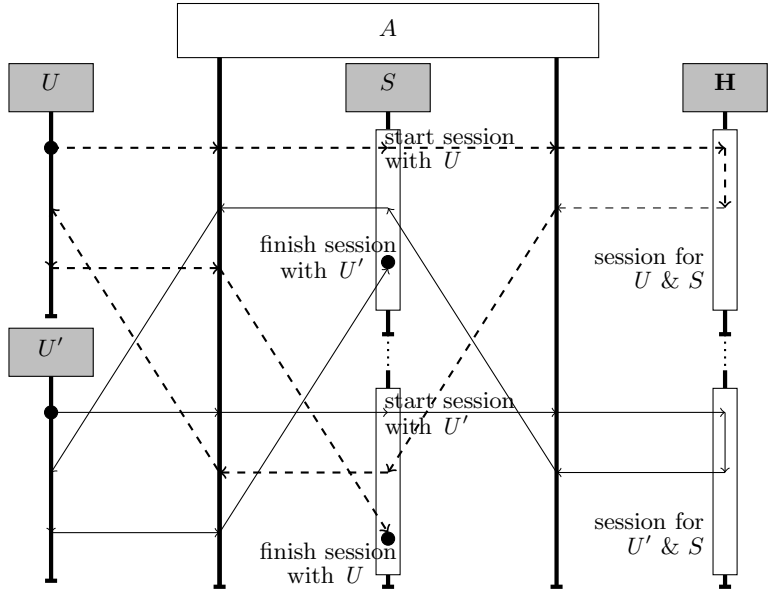

**Fig. 3.** Example of a message flow of an *outside* attack against UMTS and LTE AKA (not showing the user id request), where $U$ is authenticated to $S$ as $U'$ and $U'$ as $U$.

As a consequence, $U$ uses wireless service at the cost of $U'$ and vice versa. Furthermore, an attacker can falsely cause $U$ and $U'$ to appear physically present within certain network cells.

### 3.3 Feasibility of Real-World Attacks

In this section we consider the relevant details of the carrying protocols within the core network and discuss the feasibility of the attack presented in the previous section. The AKA messages are carried through the core network either by the Diameter protocol (for IP-based networks) or by the MAP protocol (for SS7 networks). Both carrying protocols make use of session identifier that allow network operators to match authentication data responses with the corresponding authentication data requests. The session identifier are called *Session-Id* in the Diameter specifications [37] and *Transaction ID* in the specifications of TCAP [41,42,43,44,45], which are referenced in the 3GPP MAP specification [2]. TCAP is run underneath MAP and used to handle concurrent dialogues for MAP.

The session-mixup attack of the previous section on the AKAs (with IPsec ESP, MAPsec mode2 or TCAPsec mode2) is due to either a misbinding of the authentication data response and the intended user equipment or a misbinding of the authentication data response and the corresponding authentication data request. As session identifiers bind the authentication data response and the corresponding request, the the use of such session identifiers in the carrying protocols could prevent the session mixup attack if the used session identifiers are unique and integrity protected.

In the following we consider the feasibility of the attack in the cases where the protection of the core network communication follows the network domain security specifications [5,6,7]. We stress that network domain operators are free to implement proprietary protection mechanisms for intra-domain connections.[9]

**UMTS AKA over MAP and MAPsec** When UMTS AKA is run over MAP and protected by MAPsec [5] then the session identifiers, *i.e.*, the TCAP transaction identifiers, are not protected by MAPsec and can be manipulated by an attacker. Hence, the session mixup attacks should work in practice. We note that we have not implemented the attacks yet.

**UMTS AKA over MAP and TCAPsec** When UMTS AKA is run over MAP and protected by TCAPsec [6] then the TCAP transaction identifiers are integrity protected and can generally not be manipulated by an attacker. However, the Transaction ID of TCAP, as specified in [43], may not be unique. It can be a counter of variable length from *one to four octets*. In particular, implementations of UMTS AKA over MAP with TCAP that use only one or two octet long Transaction IDs may indeed be vulnerable to the session mixup attack in practice, as a wrap-around of the Transaction ID counter could be forced by an attacker by starting a few thousand sessions. The attacker can then execute the session mixup attack for sessions with equal session identifiers. As the Transaction ID is a counter, the attacker may be able to predict which sessions eventually use the same Transaction IDs. We note that we are not aware of any real world systems that are vulnerable (due to too short session identifier). On the other hand, if the TCAP transaction identifier is long enough such that a wrap-around may not be feasible, then the TCAP transaction identifiers could be regarded as unique (in practice), and a session mixup would then not be feasible as will be shown in Section 4.

**UMTS and LTE AKA over Diameter and IPsec** When UMTS and LTE AKA are run over Diameter [37] and protected by IPsec EPS then the session identifiers are also integrity protected and can generally not be manipulated by an attacker. In terms of practical security parameters, the Session-ID of Diameter is specified in [37] to be a 64 bits long counter and is, therefore, practically unique. Again, the security proofs in Section 4 suggest that a session mixup is not feasible.

With the use of unique session identifier and authenticated encryption within the core network, the UMTS and LTE AKA protocols are more accurately described by Figure 4 than by Figure 1. In particular, the serving network $S$ in Figure 4 generates a fresh nonce $SSID$ that is included in an authentication data request. The nonce $SSID$ serves as a unique session identifier and is expected to be included in the corresponding authentication data response. Furthermore, the communication between $S$ and $H$ is protected by authenticated encryption under a key that $S$ and $H$ share.

## 4 Mechanized Analysis of the UMTS & LTE AKA

In this section, we first give an overview of the tool CryptoVerif that we used to analyze the AKA protocols and to find the flaw presented in Section 3. Then we discuss the cryptographic assumptions and modeling decisions we made. And finally we present the results we obtained.

*Notation.* A function $f : \mathbb{N} \longrightarrow \mathbb{R}_{\geq 0}$ is *negligible*, if for every positive polynomial $p$ there exists an integer $N_0$ such that for all integers $n > N_0$ it is $f(n) < 1/p(n)$. A function $g : \mathbb{N} \longrightarrow \mathbb{R}_{\geq 0}$ is *overwhelming* if $1 - g$ is a negligible function. Two sequences $\{X_\eta\}_{\eta \in \mathbb{N}}$ and $\{Y_\eta\}_{\eta \in \mathbb{N}}$ are *computationally indistinguishable* if for every polynomial-time algorithm $D$, it holds that $|\Pr[D(X_n, 1^n) = 1] - \Pr[D(Y_n, 1^n) = 1]|$ is a negligible function in $n$.

---

[9] We believe that intra-domain connections constitute the majority of connections made in the real world.

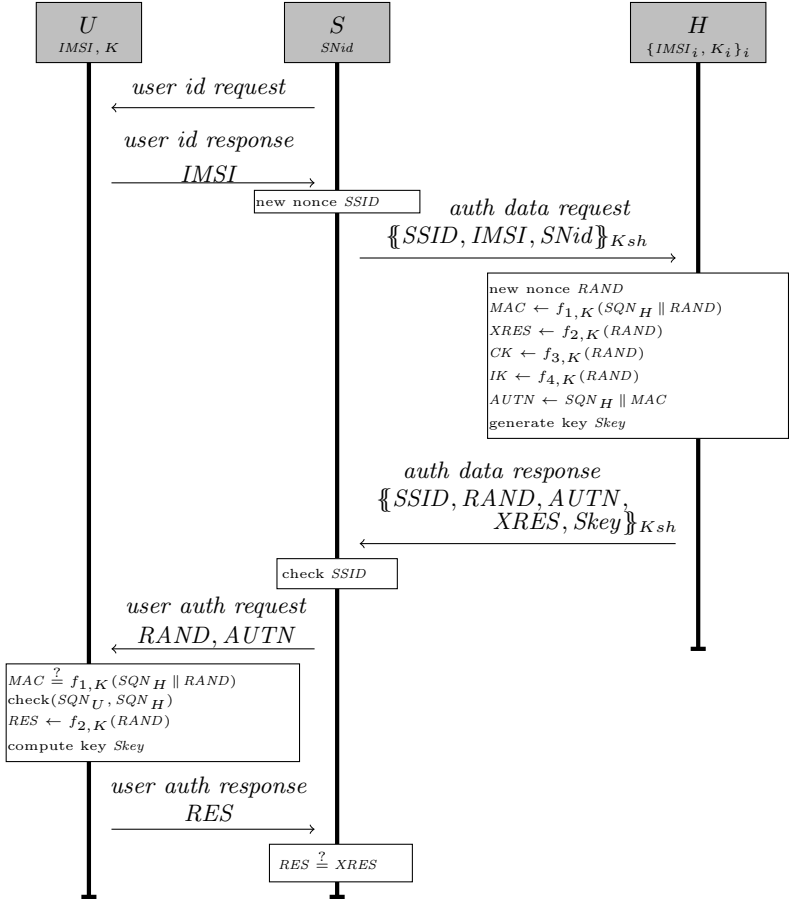

**Fig. 4.** The UMTS/LTE Authentication and Key Agreement Protocol with unique session IDs and authenticated encryption in core network. The session key in UMTS is $Skey \leftarrow CK \parallel IK$, and in LTE it is $Skey := K_{ASME} \leftarrow KDF(SQN_H \parallel CK \parallel IK \parallel SNid)$. $\{\!\{.\}\!\}_{Ksh}$ denotes authenticated encryption.

### 4.1 CryptoVerif Basics

The prover CryptoVerif [25,23,24,27] can directly prove security properties of cryptographic protocols in the computational model. Protocols are formalized in CryptoVerif using a probabilistic polynomial-time process calculus which is inspired by the pi-calculus and the calculi introduced in [46] and [51]. In this calculus, terms have a computational semantics, *i.e.* messages are bitstrings and cryptographic primitives are functions operating on bitstrings, where the lengths of bitstrings are polynomial in a security parameter $\eta$.

When analyzing protocols, CryptoVerif follows the idea of [53], *i.e.* the process calculus represents games, and CryptoVerif proofs are sequences of such games $Q_0, Q_1, \ldots, Q_n$, where the initial game $Q_0$ formalizes the protocol for which one wants to prove certain security properties. In a proof sequence, two consecutive games $Q_j$ and $Q_{j+1}$ are *observationally equivalent*, meaning that they are computationally indistinguishable for the adversary. CryptoVerif transforms one game into another by applying, *e.g.* the security definition of a cryptographic primitive or by applying syntactic transformations. For instance, if the protocol in game $Q_0$ uses a symmetric encryption scheme that is IND-CPA secure then CryptoVerif may transform game $Q_i$ to $Q_{i+1}$ by essentially replacing every occurrences of an encryption of a plaintext $m$ under a non-corrupted key with encryptions of bitstrings of zero of the same length as $m$. Therefore, all security definitions of cryptographic primitives have to be formalized as pairs of indistinguishable oracles.

In the last game in a proof sequence the desired security properties should be obvious, *i.e.* CryptoVerif uses syntactic criteria to determine whether a security property is satisfied. Given a security parameter $\eta$, CryptoVerif proofs are valid for a number of protocol sessions polynomial in $\eta$, in the presence of an active adversary. For protocols, CryptoVerif can prove secrecy properties and correspondence assertions, which can be used to formalize authentication properties.

CryptoVerif operates in two modes: a fully automatic and an interactive mode. The interactive mode requires a CryptoVerif user to input commands that indicate the main game transformations the tool should perform and the order in which they should be applied. CryptoVerif is sound with respect to the security properties it shows in a proof, but properties it cannot prove are not necessarily invalid. Occasionally, one needs to conclude a proof manually by inspecting the last game that one obtained using either the fully automatic or the interactive mode of CryptoVerif. A more detailed description of CryptoVerif and its process calculus is given in [25].

**CryptoVerif Language** In CryptoVerif, a term $M$ that represent computations on bitstrings can be constructed according to the following grammar:

$$
\begin{array}{lll}
M ::= & & \text{term} \\
& i & \text{replication index} \\
& x[M_1, \ldots, M_m] & \text{variable access} \\
& f(M_1, \ldots, M_m) & \text{function application,}
\end{array}
$$

where replication index $i$ is used to distinguish several copies of a replicated process, and the variable access $x[M_1, \ldots, M_m]$ returns the value of $x$ for the indices $M_1, \ldots, M_m$.

There are two kinds of processes in the calculus: input processes and output processes. These can be generated according to the grammar depicted in Figure 5.

$$
\begin{array}{lll}
Q ::= & & \text{input process} \\
& 0 & \text{nil} \\
& Q \mid Q' & \text{parallel composition} \\
& !^{i \leq N} Q & \text{replication } N \text{ times} \\
& \mathsf{newChannel}\ c; Q & \text{channel restriction} \\
& c[M_1, \ldots, M_l](x_1[\tilde{i}] : T_1, \ldots, x_k[\tilde{i}] : T_k); P & \text{input} \\
& & \\
P ::= & & \text{output process} \\
& \overline{c[M_1, \ldots, M_l]}\langle M'_1, \ldots, M'_k \rangle; Q & \text{output} \\
& \mathsf{new}\ x[i_1, \ldots, i_m] : T; P & \text{random number} \\
& \mathsf{let}\ x[i_1, \ldots, i_m] : T = M\ \mathsf{in}\ P & \text{assignment} \\
& \mathsf{if\ defined}(M_1, \ldots, M_l) \wedge M\ \mathsf{then}\ P\ \mathsf{else}\ P' & \text{conditional} \\
& \mathsf{find}\ (\bigoplus_{j=1}^m u_{j1}[\tilde{i}] \leq n_{j1}, \ldots, u_{jm_j}[\tilde{i}] \leq n_{jm_j} & \\
& \ \ \mathsf{suchthat\ defined}(M_{j1}, \ldots, M_{jl_j}) \wedge M_j\ \mathsf{then}\ P_j) & \\
& \ \ \mathsf{else}\ P' & \text{array lookup} \\
& \mathsf{event}\ e(M_1, \ldots, M_m); P & \text{event,}
\end{array}
$$

**Fig. 5.** Syntax of CryptoVerif's process calculus

The nil process does nothing, $Q \mid Q'$ represents parallel execution of $Q$ and $Q'$, the process $!^{i \leq N} Q$ represents parallel execution of $N$ copies of $Q$, which are indexed by $i$, and $\mathsf{newChannel}\ c; Q$ creates a new private channel and executes $Q$. When process $\overline{c[M_1, \ldots, M_l]}\langle N_1, \ldots, N_k \rangle; Q$, *e.g.* representing a message being sent over the network, is executed then an input $c[M'_1, \ldots, M'_l](x_1[\tilde{i}] : T_1, \ldots, x_k[\tilde{i}] : T_k); P$ (where $T$ represents types used in the calculus) is searched that is ready to receive the sent message, *i.e.* the instantiations of the $M'_i$ should be equal to the bitstring representation of the $M_i$ – if none exists then the process is blocked, if there are several then one is chosen at uniform random, and then for $j = 1, \ldots, k$, each $N_j$ is essentially stored in $x_j[i]$. Afterwards $P$ is executed and input process $Q$ becomes one of the available input processes. The process new $x[i_1, \ldots, i_m] : T; P$ chooses uniformly at random a number in the set of bitstrings corresponding to $T$ (*e.g.* a set of random seeds needed for probabilistic encryption), let $x[i_1, \ldots, i_m] : T = M$ in $P$ stores the bitstring instantiation of $M$, which must be of type $T$, in $x[i_1, \ldots, i_m]$ and proceeds with $P$. The process if defined$(M_1, \ldots, M_l) \wedge M$ then $P$ else $P'$ executes $P$ if $M_1, \ldots, M_l$ have been defined and the interpretation of $M$ (*e.g.*, an equality test on bitstrings) holds, otherwise it executes $P'$. The process find $(\bigoplus_{j=1}^{m} u_{j1}[\tilde{i}] \leq n_{j1}, \ldots, u_{jm_j}[\tilde{i}] \leq n_{jm_j}$ suchthat defined$(M_{j1}, \ldots, M_{jl_j}) \wedge M_j$ then $P_j$) else $P'$, where $\tilde{i}$ is a tuple $i_1, \ldots, i_{m'}$, tries to find a branch of the execution indexed by $1 \leq j \leq m$ such that there exist indices $u_{j1}[\tilde{i}] \leq n_{j1}, \ldots, u_{jm_j}[\tilde{i}] \leq n_{jl_j}$ for which $M_{j1}, \ldots, M_{jl_j}$ have been defined and the computational interpretation of $M_j$ holds, and in that case the process continues by executing $P_j$, otherwise it executes $P'$. The process event $e(M_1, \ldots, M_m); P$ executes the event $e(M_1, \ldots, M_m)$, then executes $P$. Executing an event does not change the state of the system, but events are used in the specification of authentication properties.

As an example, we formalize in CryptoVerif the process of the user equipment in the case of LTE AKA. Additional formalizations of processes for LTE AKA and LTE AKA+1 can be found in the Appendix A. Note that we do not model the first message by the serving network in Figure

$$
\begin{aligned}
Q_U = &\,!^{i_U \leq N_U}\, c_1[i_U](hostS : sn); \\
&\overline{c_2[i_U]}\langle U, hostS \rangle; \\
&c_{11}[i_C](RAND'' : nonce, = SQN, MAC'' : prpcblocksize); \\
&\text{let } MAC''' = \text{prpcenc}(\text{concat}_8(const_1, SQN, RAND), prpKuh) \text{ in} \\
&\text{if } MAC''' = MAC'' \text{ then} \\
&\text{let } RES'' = \text{prpcenc}(\text{concat}_9(const_2, RAND), prpKuh) \text{ in} \\
&\text{let } CK'' = \text{prpcenc}(\text{concat}_9(const_3, RAND), prpKuh) \text{ in} \\
&\text{let } IK'' = \text{prpcenc}(\text{concat}_9(const_4, RAND), prpKuh) \text{ in} \\
&\text{let } K''_{ASME} = \text{prf}(prfKuh, \text{concat}_3(SQN, CK'', IK'', hostS)) \\
&\text{event } partAuthU(RAND'', RES''); \\
&\overline{c_{12}[i_U]}\langle RES'' \rangle; \\
&c_{finish}[i_U](); \\
&\text{if } hostS = S \text{ then} \\
&( \\
&\quad \text{event } endAuthU2(hostS, RAND'', K''_{ASME}); \\
&\quad \text{let } keyU : \text{mkeyseed} = K''_{ASME} \\
&) \\
&\text{else } \overline{c_{end_1}[i_U]}\langle K''_{ASME} \rangle.
\end{aligned}
$$

**Fig. 6.** CryptoVerif formalization of user equipment actions in LTE AKA

1, *i.e.* the user id request. First process $Q_U$ obtains the name of the $S$ to interact in an AKA run with, then $Q_U$ sends his own name (representing his IMSI) and the name $S$ of a serving network out. When $Q_U$ receives a nonce $RAND''$, the sequence number $SQN$ equal to his own, and $MAC''$ that is of type *prpcblocksize*, *i.e.* in the range of the PRP block cipher prpcenc (see Section 4.2 for the cryptographic assumptions made), then $Q_U$ re-computes and checks the $MAC''$ using the long-term key $prpKuh$ shared with trusted third party $H$ and a public constant $const_1$. Afterwards $Q_U$ computes $RES''$, $CK''$ and $IK''$, where $Q_U$ uses the same key $prpKuh$ but different constants $const_2, const_3, const_4$. The patterns concat$_8$ and concat$_9$ are used to wrap messages of different formats so that they can be taken as the first argument of the PRP encryption function. Then $Q_U$ computes the session key $K''_{ASME}$ with a pseudorandom function and a key for that purpose that it shares with $H$. Before $Q_U$ sends out $RES''$, it records some of the messages seen in this run in the event $partAuthU$, which should then correspond to messages recorded in events of the process for honest $S$ in order to obtain authentication guarantees. Finally, if $Q_U$ indeed communicated with $S$ then records the name $S$ in the event $endAuthU2$ and stores $K''_{ASME}$ in $keyU$, but if $hostS$ was not honest $S$ (but the adversary) then $Q_U$ simply publishes the key.

**Authentication and Secrecy using CryptoVerif** Authentication in CryptoVerif is modeled by correspondence properties [24]. Events $e(M_1, \ldots, M_m)$ are used in order to record that a certain program point has been reached, with certain values of $M_1, \ldots, M_m$, and the correspondence properties are properties of the form "if some event has been executed, then some other events also have been executed, with overwhelming probability".

> A process $Q$ *satisfies the correspondence* $\mathsf{event}(e(M_1, \ldots, M_m)) \Rightarrow \bigwedge_{i=1}^{k} \mathsf{event}(e_i(M_{i1}, \ldots, M_{im_i}))$ if and only if, with overwhelming probability, for all values of the variables in $M_1, \ldots, M_m$, if the event $e(M_1, \ldots, M_m)$ has been executed, then the events $e_i(M_{i1}, \ldots, M_{im_i})$ for $i \leq k$ have also been executed for some values of the variables of $M_{ij}$ ($i \leq k$, $j \leq m_i$) not in $M_1, \ldots, M_m$.

Note that CryptoVerif can also show stronger, *injective* correspondence, where for $\mathsf{event}(e(M_1, \ldots, M_m)) \Rightarrow \bigwedge_{i=1}^{k} \mathsf{event}(e_i(M_{i1}, \ldots, M_{im_i}))$ the executions of the events $e_i(M_{i1}, \ldots, M_{im_i})$ are distinct. However, we are not interested in injective correspondences for UMTS and LTE AKA, as they simply do not hold given our modeling of the protocol where there is little replay protection. The formal definitions of correspondences in CryptoVerif can be found in [24].

A variable is considered secret when the adversary has no information on it, that is, the adversary cannot distinguish it from a random number.

> A process $Q$ *preserves the one-session secrecy of $x$* when, with overwhelming probability, the adversary interacting with $Q$ cannot distinguish any element of the array $x$ from a uniformly distributed random number by a single test query. The test query returns either the desired element of $x$ or a freshly generated random number, and the adversary has to distinguish between these two situations.

Note that this notion of secrecy corresponds to the standard notion of key indistinguishability; but it does not guarantee that the random numbers in $x$ are independent. CryptoVerif can show a stronger notion of secrecy, in which the adversary can perform several test queries, and which therefore corresponds to the "real-or-random" definition of security [13]. However, this stronger notion is not satisfied by UMTS and LTE AKA the way we model them. In particular, as replays are not immediately rejected by the protocols in our model, an adversary can force several protocol sessions to agree on the same key; if the adversary is allowed to perform several test queries then he can distinguish a challenge key with overwhelming probability. The formal definitions of secrecy in CryptoVerif can be found in [25].

### 4.2 Cryptographic Assumptions

As CryptoVerif operates with computational semantics, we need to specify the assumptions we make for the cryptographic primitives used in UMTS and LTE AKA. The algorithms for $f_1, \ldots, f_4$ (see Figure 1) are not mandated by the 3GPP's specifications, but can be chosen freely by the home network operator. Indeed $f_1, \ldots, f_4$ are only needed in the USIM of the user equipment, and in the AuC server in the home network but are not used by the serving network $S$. These functions are rather vaguely described in [9]: $f_1$ and $f_2$ are 'message authentication functions' and $f_3, f_4, f_5$ are 'key generating functions'. However, 3GPP specifies the sample algorithms set MILENAGE [10]. Motivated by MILENAGE, we assume that the $f_1, \ldots, f_4$ are all based on a single pseudo-random permutation block cipher[10]. Notice that we assume that no *anonymity key* is used, *i.e.* $f_5 \equiv 0$, which is an option in [9,4]. As in MILENAGE, all $f_i$ in the same run will use the same long-term key (shared between $U$ and $H$) in our model but each $f_i$ also takes as input a constant $c_i$. We further assume that each $S$ shares with $H$ a long-term symmetric encryption key and a long-term message authentication key; which corresponds to the requirement that IPsec and MAPsec must support pre-shared keys and also to our assumption that IPsec and MAPsec security associations are static. Moreover, we assume that these keys are used to protect the connection between a $S$ and $H$ through an encrypt-then-mac scheme, where the encryption is IND-CPA secure and the message authentication code is WUF-CMA secure (this implies INT-PTXT [19]). There is no ordering on terms by size and no arithmetics in CryptoVerif, so we cannot implement any checks on the sequence number $SQN$ as intended by AKA; in our model the sequence number is a constant and the user only checks for equality. In particular, that means that the protocol in our model lacks replay-attack protection. In addition, in the case of LTE AKA, we assume that the key derivation function is a pseudo-random function which outputs a key seed to generate a message authentication key. This key seed is then used to generate the session key $K_{ASME}$. This way of indirectly generating $K_{ASME}$ is due to the fact that in cryptography, security definitions for a message authentication code (but also for encryption schemes etc.) only hold for keys that are created by the accompanying key generation algorithm. All cryptographic primitives that we use are already modeled and ready-to-use in CryptoVerif.

---

[10] in CryptoVerif the PRP block cipher is actually modeled rather like a PRF; this is justified by the PRF/PRP switching lemma, *e.g.* [20]

### 4.3 Results

In this section, we present results that we have obtained with CryptoVerif (version 1.16) when analyzing the UMTS and LTE AKA as depicted in Figure 4 under the cryptographic assumption stated in Section 4.2. In all the results below we assume that the home network $H$ is a trusted third party which always acts honestly. Excerpts of the CryptoVerif scripts used for the analysis of LTE AKA and LTE AKA+1 (as defined in Section 2.2) are given by Figure 6 and by Figures 7 – 13, which can be found in the appendix A. The corresponding input scripts for UMTS AKA are easily derivable from the presented scripts for LTE AKA; the main difference lies in the session key (cf. Figure 1), *i.e.* for UMTS AKA, the process for $H$ does not generate $K_{ASME}$ and instead of $K_{ASME}$ the pair of keys $(CK, IK)$ is used in the sent messages and events.[11]

**Results for UMTS AKA** For UMTS we can prove the following.

**Theorem 1 (Authentication of core network to User Equipment).** *In the UMTS AKA, if there is an instance of*

- *an honest user $U$ receiving a value $RAND''$ as nonce in a user authentication request and generates with it and the key shared with home network $H$ a response $RES''$*

*then, with overwhelming probability, there is an instance of*

- *the home network $H$ completing a run of the UMTS AKA for serving network $S$ and user $U$*
- *in which $H$ generated a nonce $RAND$ and an expected response $XRES$, where $RAND$ equals $RAND''$ and $XRES$ equals $RES''$.*

Intuitively, this theorem implies that if a users completes a run of the UMTS AKA protocol then the users can be certain that the received nonce was generated by the trusted home network (although this does not give the user any guarantee that the serving network it communicates with is acting honestly).

**Theorem 2 (Entity Authentication of User Equipment to Serving Network).** *In the UMTS AKA, if there is an instance of*

- *an honest serving network $S$ completing a run of the UMTS AKA with honest user equipment $U$ and home network $H$*

---

[11] Please contact the authors of this work directly to obtain the full CryptoVerif input scripts for UMTS AKA, LTE AKA or LTE AKA+1.

- in which $S$ received a value $RAND'$ as nonce and a value $XRES'$ as expected response from $H$ in a authentication data response
- and in which $S$ received a value $RES'$ that equals $XRES'$ in a user authentication response

then, with overwhelming probability, there is an instance of

- $H$ completing a data authentication transfer with $S$
- in which $H$ generated a nonce $RAND$ and an expected response $XRES$ for the use between $S$ and $U$, where $RAND'$ equals $RAND$ and $XRES'$ equals $XRES$

and an instance of

- $U$ completing a run of the UMTS AKA
- in which $U$ received a value $RAND''$ as nonce that is equal to $RAND'$
- and in which $U$ sent a response $RES''$ that equals $XRES'$.

Intuitively, this theorem implies that if serving network $S$ completes a run of UMTS AKA for user $U$ then $U$ must indeed have been involved in a run of the UMTS AKA in which it generated matching values.

**Theorem 3 (Key Secrecy in UMTS AKA).** *Let $Q_{UMTS}$ be the game in CryptoVerifs process calculus formalizing the UMTS AKA. Furthermore, let keyS1 and keyS2 denote in $Q_{UMTS}$ the confidentiality key CK and, respectively, the integrity key IK that are received by an honest serving network from the home network and generated by the home network for the use between the serving network and an honest user. Then $Q_{UMTS}$ preserve the one-session secrecy of keyS1 and keyS2.*

This theorem states that the exchanged keys $CK$, $IK$ that an honest serving network $S$ holds after completing a UMTS AKA run with honest user $U$ are cryptographically secret (in the sense of 1-session secrecy). It is already known that UMTS AKA does not offer entity authentication of a serving network $S$ to user equipment $U$ (which, *e.g.*, allows an attacker to set up false base stations; see [55]). Therefore, with respect to secrecy of exchanged session keys, we can only hope for showing that the key that $S$ holds at the end of a session with an honest $U$ is secure. The process of user $U$ cannot distinguish cases where it agreed on a session key with honest serving network $S$ from cases where the session key is agreed on with a dishonest serving network $S'$. Therefore $U$'s session key is not secret.

*Proof (for Theorems 1, 2 & 3).* In the CryptoVerif process formalizing the UMTS AKA, when the process for honest $S$ completes a run of the AKA protocol it executes an event $endAuthS(h, RAND', XRES')$ that contains the name $h$ of the user equipment that supposedly participated in the protocol run, and a nonce $RAND'$ and an expected response $XRES'$ that were protected under a encrypt-then-mac scheme with key that is shared with $H$. The event is executed after $S$ received a response that equals $XRES'$. When the process for $H$ completes a run of the authentication data transfer it executes an event $endAuthH(h2, h1, RAND, XRES)$ that contains the name of a registered serving network $h1$ from whom it received a request, the name $h2$ of a registered user equipment $U$, a random nonce $RAND$ generated by $H$, and an expected response $XRES$ generated by $H$ using a key shared between $H$ and $h2$. Furthermore, it executes an event event $endAuthHU(h2, RAND, XRES)$. When the process for honest $U$ completes a run of the AKA protocol then it executes an event $partAuthU(RAND'', RES'')$ that contains a nonce $RAND''$ that $U$ received together with a valid $MAC$ produced with a key $k_{uh}$ shared between $U$ and $H$, and that contains a response $RES''$ generated by $U$ over $RAND''$ using the key $k_{uh}$. We can show with CryptoVerif the correspondence queries

$$\text{event } partAuthU(x, y) \Rightarrow \text{event } endAuthHU(U, x, y)$$
$$\text{event } endAuthS(U, x, y) \Rightarrow \text{event } endAuthH(U, S, x, y)$$
$$\text{event } endAuthS(U, x, y) \Rightarrow \text{event } partAuthU(x, y).$$

The first of these queries captures the statement of Theorem 1.

With respect to key secrecy: In $Q_{UMTS}$, when the process for the honest serving network completes a run in the UMTS AKA with $H$ and an honest user $U$, then it stores the session keys received from $H$ in *keyS1* and *keyS2*, respectively. CryptoVerif can prove the one-session secrecy queries `secret1` *keyS1* and `secret1` *keyS2* automatically.

The proof for Theorems 1, 2 & 3 takes 44 game transformations for CryptoVerif using the commands

1. `auto`
2. `SArename RAND_255;`
3. `SArename @22_r2_272;`
4. `SArename @22_r2_270;`
5. `move array @22_r2_493;`
6. `success.`

□

**Results for LTE AKA** For LTE AKA we can prove the following.

**Theorem 4 (Authentication of core network to User Equipment).** *In the LTE AKA, if there is an instance of*

- *an honest user $U$ receiving a value $RAND''$ as nonce in a user authentication request and generates with it and the key shared with home network $H$ a response $RES''$*

*then, with overwhelming probability, there is an instance of*

- *the home network $H$ completing a run of the LTE AKA for serving network $S$ and user $U$*
- *in which $H$ generated a nonce $RAND$ and an expected response $XRES$, where $RAND$ equals $RAND''$ and $XRES$ equals $RES''$.*

Again, as for UMTS AKA, the LTE AKA itself does not offer authentication of the serving network to the user, but we can only show that a serving network that a user equipment communicates with has been authorized by the home network.

**Theorem 5 (Entity Authentication of User Equipment to Serving Network).** *In the LTE AKA, if there is an instance of*

- *an honest serving network $S$ completing a run of the LTE AKA with honest user equipment $U$ and home network $H$*
- *in which $S$ received a value $RAND'$ as nonce and a value $XRES'$ as expected response from $H$ in a authentication data response*
- *and in which $S$ received a value $RES'$ as response in a user authentication response that equals $XRES'$*

*then, with overwhelming probability, there is an instance of*

- *$H$ completing a data authentication transfer with $S$*
- *in which $H$ generated a nonce $RAND$ and an expected response $XRES$ for the use between $S$ and $U$, where $RAND'$ equals $RAND$ and $XRES'$ equals $XRES$*

*and an instance of*

- *$U$ completing a run of the LTE AKA*
- *in which $U$ received a value $RAND''$ as nonce in an user authentication request that is equal to $RAND'$*
- *and in which $U$ sent a response $RES''$ that equals $XRES'$.*

As for UMTS AKA, the LTE AKA itself does not offer authentication of the serving network to the user, therefore we cannot show key secrecy for the session key held by the honest user. But we can again show secrecy for the session key of the serving network.

**Theorem 6 (Key Secrecy in LTE AKA).** *Let $Q_{LTE}$ be the game in CryptoVerifs process calculus formalizing the LTE AKA. Furthermore, let keyS denote in $Q_{LTE}$ the session key $K_{ASME}$ received by an honest serving network from the home network and generated by the home network for the use between the serving network and an honest user. Then $Q_{LTE}$ preserve the one-session secrecy of keyS.*

*Proof (for Theorems 4, 5 & 6).* The proof is analogous to the proof of Theorems 1, 2 & 3, using the commands

1. `auto`
2. `SArename RAND_275;`
3. `SArename @22_r2_292;`
4. `SArename @22_r2_290;`
5. `move array @22_r2_2150;`
6. `success`.

CryptoVerif needs 57 transformations. □

Unlike UMTS AKA, the LTE is designed to offer also authentication of $S$ to $U$. For this reason the session key $K_{ASME}$ is generated using $S$'s identity. Nevertheless, the authentication guarantee is not explicitly established in LTE AKA itself but should hold after the next message sent by $S$ to $U$, *i.e.* the first message of the exchange called *NAS Security Mode Command Procedure* (NAS SMC) in [9]. This exchange uses $K_{ASME}$ [12] as key for a message authentication code. Recall from Section 2.2, that we denote the LTE AKA with the additional first message of the NAS SMC by LTE AKA+1. For LTE AKA+1, we can show, in addition to Theorem 4 and 5, the following.

**Theorem 7 (Entity Authentication of Serving Network to user equipment).** *In the LTE AKA+1, if there is an instance of*

– *an honest user equipment $U$ completing a run of the LTE AKA+1 with honest serving network $S$ and home network $H$*
– *in which $U$ received a value $RAND''$ as nonce in an user authentication request and derived a key seed $K''_{ASME}$*
– *and in which $U$ verified a valid message authentication code on a received message using the key generated from $K''_{ASME}$*

*then, with overwhelming probability, there is an instance of*

– *$H$ completing a data authentication transfer with $S$*
– *in which $H$ generated a nonce $RAND$ and derived a key seed $K_{ASME}$ for the use between $S$ and $U$, where $RAND''$ equals $RAND$ and $K''_{ASME}$ equals $K_{ASME}$*

*and an instance of*

– *$S$ completing a run of the LTE AKA+1*
– *in which $S$ received a value $RAND'$ as nonce from $H$ that is equal to $RAND''$ and derived a key seed $K'_{ASME}$ that equals $K''_{ASME}$*
– *and in which $S$ sent out a NAS security mode command message with a message authentication code under the key generated from $K'_{ASME}$.*

*Proof (of Theorem 7).* In the CryptoVerif process that formalizes the LTE AKA+1, when the process for honest user equipment $U$ completes a run of the AKA+1 protocol, it executes an event $endAuthU1(hostS, RAND'', K''_{ASME})$ that contains the name $hostS$ of the serving network that supposedly participated in the same run, a nonce $RAND''$ that $U$ received together with a valid $MAC$ produced with a key $k_{uh}$ shared between $U$ and $H$, and a mac key seed $K''_{ASME}$ that $U$ derived from inputs including $RAND''$, the name $hostS$ and a key shared between $U$ and $H$.

---
[12] in our model a key generated from $K_{ASME}$

When the process for $H$ completes a run of the authentication data transfer it executes an event $endAuthH1(h2, h1, RAND, K_{ASME})$ that contains the name of a registered serving network $h1$ from whom it received a request, the name $h2$ of a registered user equipment $U$, a random nonce $RAND$ generated by $H$, and a key seed $K_{ASME}$ derived by $H$ from inputs including $RAND$, the name $h1$ and a key shared between $H$ and $h2$. We can show with CryptoVerif the correspondence queries

$$\text{event } endAuthU1(S, x, y) \Rightarrow \text{event } endAuthS1(U, x, y)$$
$$\text{event } endAuthU1(S, x, y) \Rightarrow \text{event } endAuthH1(U, S, x, y).$$

We can show the queries with CryptooVerif automatically in 60 game transformations. □

*Remark 1.* Although LTE AKA+1 satisfies authentication of $S$ to $U$, the session key that $U$ holds at the end of a completed session is still not computationally secret. The reason in this case is the following: (a key derived from) this session key is used during the session for generating a message authentication code. This allows an attacker to distinguish the key easily. Given a challenge key, the attacker just tries to verify this message authentication code with the challenge key. If the verification is successful, then the attacker guesses that the challenge key equals the session key, else the challenge key must be random key. Thus, the attacker has a overwhelming success probability in distinguishing $U$'s session key. Moreover, for the same reason any session key that honest serving network $S$ holds at the end of a completed run does also not satisfy one-session secrecy, *i.e.* the result from Theorem 6 does not hold for LTE AKA+1.

## 5  Conclusions and Future Work

We present the first computational security analysis of the Authentication and Key Agreement of UMTS and LTE, where we consider entity authentication and key secrecy properties while taking into account the carrying protocols within the core network. The analysis uncovers a flaw in the specifications of UMTS and LTE AKA and the network domain security with rather serious consequences. An outside attacker can defeat entity authentication of the user equipments to the serving networks. And an inside attacker can authenticate as another honest subscriber to a serving network and use the wireless services on his behalf. Previous published analyses of UMTS AKA were working in a purely symbolic model and did not catch the attacks because they generally made too strong assumptions on the connection between serving network and home network.

The attacks can be prevented if the carrying protocols use unique integrity-protected session identifiers in the communication between serving and home network. In practice, UMTS AKA is vulnerable if it is run over MAP and MAPsec, as the session identifiers are not integrity-protected, while UMTS and LTE AKA run over the Diameter protocol and IPsec are secure. We use the tool CryptoVerif to verify entity authentication properties and key secrecy properties with respect to the computational model for the UMTS and LTE AKA protocols when unique session IDs and authenticated encryption in the core network are used.

We question whether it is prudent practice to make the security of the UMTS/LTE AKA protocol (or, in fact, any other cryptographic protocol) reliant on the carrying protocols and their protection without considering the security of the combined protocol. Instead we believe that it would be much more desirable to directly consider the combination of AKA with carrying protocol and its protection. The UMTS/LTE AKA protocols should ideally be strengthened by making the binding of $H$'s authentication data response for an intended $U$ explicit in the AKA specifications and by explicitly requiring that authenticated encryption is used to protect the connection between serving and home network. Moreover, the specifications [4] and [9] of UMTS/LTE AKA need to be revised so that the desired security properties on all connections are explicitly stated. This is also important for intra-domain connections, where operators are free to implement proprietary solutions and may currently miss to implement or sufficiently protect mechnisms for the serving network to correctly match the authentication data responses, even if their implementation is guided by the specifications of the security architecture [4,9].

We are interested in further exploring to what extend real-world systems are vulnerable to our attack, which, *e.g.*, also depends on how widely MAPsec or TCAPsec with short transaction identifiers are deployed in practice. We would also like to expand our analysis to scenarios of the protocol execution that are not covered in the present work, *e.g.* the scenarios that are related to the use of TMSIs, and include the usage of the sequence number in order to verify stronger authentication properties than entity authentication. Moreover, it would be interesting to verify under which conditions the specified MILENAGE algorithms and the key derivation algorithm satisfy the computational assumptions made in this work.

**Acknowledgements** We thank Valtteri Niemi and Steve Babbage for helpful discussions on the feasibility of the session-mixup attack. And we thank Bruno Blanchet for helping to automate the proofs for entity authentication of the user to the serving network.

## A  LTE AKA in CryptoVerif

We present, in Figures 7 – 13, additional parts of the input scripts to CryptoVerif used in our analysis of LTE AKA and LTE AKA+1, which we believe are sufficient to understand the modeling of the protocol and the explanations of the proofs in this paper. Please see the CryptoVerif manual, which ships with the tool downloadable at `http://www.cryptoverif.ens.fr/`, for further detail on the syntax of the input language.

**Notation** *ue, sn* and *hn* denote the types of, respectively, user equipments, serving networks and home networks. *mkeyseed, mkey* and *macs* denote the types of, respectively, inputs to the key generator, keys, and outputs of the message authentication code (which is used for protection of the communication between serving network and home network). Similarly, *keyseed, key* and *seed* denote the types of, respectively, inputs to the key generator, keys and random seeds used by the probabilistic encryption scheme (which ist used for protection of the communication between serving network and home network); the cleartexts and ciphertexts of the encryption scheme are of type *bitstring*. *prfkeyseed, prfkey* and *prft* denote the types of, respectively, inputs to the key generator, keys and outputs of the pseudo-random function (which is used as key derivation function). *prpckeyseed, prpckey* and *prpcblocksize* denote the types of, respectively, inputs to the key generator, keys and outputs of the pseudo-random permutation block cipher (which is used for the functions $f_1, \ldots, f_4$). And *econst* denotes the type of constants that are used as input to the prp block cipher in order to distinguish its usage as $f_1$, ..., or $f_4$.

kgen, enc and dec comprise the encryption scheme used to protect messages between serving network and home network. $i_\perp$ is the natural injection from the set of cleartexts to *bitstring* $\cup \perp$ (needed as decryption may fail). The patterns concat$_j$ for $j \in \{1, \ldots, 9\}$ are used to wrap messages of different formats so that they can be taken as input of a function, *e.g.* the encryption function. mkgen, mac and check comprise the message authentication code used to protect messages between serving network and home network. prpcenc and prpckgen denote the pseudo-random permutation block cipher and its key generation function used to model the functions $f_1, \ldots, f_4$. prf and prfkgen denote the pseudo-random function and its key generation function used to model the key derivation function.

Otherwise the scripts follow the syntax as presented in Figure 5. Please contemplate the full input scripts and CryptoVerif's manual for the exact type declarations and exact function declaration/expansion. The main process is given in Figure 13, which is equal for LTE AKA and LTE AKA+1 up to the different sub processes for user equipment and serving network. Before executing the processes for user equipment, serving network and home network in parallel, the main process first generates values, *e.g.* key material, on which the adversary has *a priori* no information. The two processes presented in Figures 10 and 11 allow the adversary to register with the home network as, respectively, serving network or user equipment (as needed for inside attacks). They are used by the home network $H$ to look up long-term keys shared between $H$ and a serving network or a user equipment.

$$\begin{aligned}
Q_S = {}& !^{i_S \leq N_S} c_3[i_S](h : ue, hostS' : sn); \\
& \text{if } hostS' = S \text{ then} \\
& \text{new } r_1 : seed; \\
& \text{let } e_1 = \text{enc}(\text{concat}_2(h, S), Ksh, r1) \text{ in} \\
& \overline{c_4[i_S]}\langle S, e_1, \text{mac}(e_1, mKsh)\rangle; \\
& c_9[i_S](e'_4 : bitstring, mac'_4 : macs); \\
& \text{if check}(e'_4, mKsh, mac'_4) \text{ then} \\
& \text{let } \text{i}_\perp(\text{concat}_7(= h, K'_{ASME}, RAND', XRES', SQN', MAC')) = \text{dec}(e'_4, Ksh) \text{ in} \\
& \overline{c_{10}[i_S]}\langle RAND', SQN', MAC'\rangle; \\
& c_{13}[i_S](RES' : prpcblocksize); \\
& \text{if } XRES' = RES' \text{ then} \\
& \text{event } endAuthS(h, RAND', RES'); \\
& \text{if } h = U \text{ then} \\
& ( \\
& \text{let } keyS : mkeyseed = K'_{ASME} \\
& ) \\
& \text{else } \overline{c_{end_2}[i_S]}\langle K'_{ASME}\rangle.
\end{aligned}$$

**Fig. 7.** CryptoVerif formalization of serving network (eNB/MME) actions in LTE AKA

## B  GSM Authentication Protocol

The GSM Subscriber Identity Authentication protocol is depicted in Figure 14 and can be regarded as the 2G predecessor of the UMTS AKA protocol. When the home network $H$ receives a security information request for $U$, then $H$ looks up the individual subscriber authentication key $K_0$ that it shares with $U$ and generates the authentication vector response by generating a random $RAND$ and by computing the signed response $SRES$. The *authentication algorithm* $A_3$ and the *ciphering key generating algorithm* $A_8$ fulfill similar tasks as, respectively, $f_2$ and $f_3/f_4/KDF$ for UMTS/LTE AKA.

The GSM SIA is specified in [12,11] and depicted in Figure 14 in Appendix B. This protocol suffers from the same underspecification as UMTS and LTE AKA: there is no proper binding of the response from the home network (called *Authentication Vector Response*) to the corresponding request or user. Therefore, both attacks of Figures 3 and 2 could also be deployed against GSM SIA. However, the case of GSM SIA is different from the case UMTS/LTE AKA. In the GSM

$$Q_U^+ = !^{i_U \leq N_U} c_1[i_U](hostS : sn);$$
$$\overline{c_2[i_U]}\langle U, hostS \rangle;$$
$$c_{11}[i_C](RAND'' : nonce, = SQN, MAC'' : blocksize);$$
$$\text{let } MAC''' = \text{prpcenc}(\text{concat}_1(const_1, SQN, RAND), prpKuh) \text{ in}$$
$$\text{if } MAC''' = MAC'' \text{ then}$$
$$\text{let } RES'' = \text{prpcenc}(\text{concat}_2(const_2, RAND), prpKuh) \text{ in}$$
$$\text{let } CK'' = \text{prpcenc}(\text{concat}_2(const_3, RAND), prpKuh) \text{ in}$$
$$\text{let } IK'' = \text{prpcenc}(\text{concat}_2(const_4, RAND), prpKuh) \text{ in}$$
$$\text{let } K''_{ASME} = \text{prf}(prfKuh, \text{concat}_3(SQN, CK'', IK'', hostS))$$
$$\text{event } partAuthU(RAND'', RES'');$$
$$\overline{c_{12}[i_U]}\langle RES'' \rangle;$$
$$c_{15}[i_U](= NASsma, mac_6 : macs);$$
$$\text{let } mKNAS'' = \text{mkgen}(K''_{ASME}) \text{ in}$$
$$\text{if check}(NASsma, mKNAS'', mac_6) \text{ then}$$
$$\text{if } hostS = S \text{ then}$$
$$\text{event } endAuthU1(hostS, RAND'', K''_{ASME}).$$

**Fig. 8.** CryptoVerif formalization of user equipment actions in LTE AKA+1, where $NASsma$ is considered to be a non-secret constant

$$Q_S^+ = !^{i_S \leq N_S} c_3[i_S](h : ue, hostS' : sn);$$
$$\text{if } hostS' = S \text{ then}$$
$$\text{new } r_1 : seed;$$
$$\text{let } e_1 = \text{enc}(\text{concat}_2(h, S), Ksh, r1) \text{ in}$$
$$\overline{c_4[i_S]}\langle S, e_1, \text{mac}(e_1, mKsh) \rangle;$$
$$c_9[i_S](e'_4 : bitstring, mac'_4 : macs);$$
$$\text{if check}(e'_4, mKsh, mac'_4) \text{ then}$$
$$\text{let } i_\perp(\text{concat}_7(= h, K'_{ASME}, RAND', XRES', SQN', MAC')) = \text{dec}(e'_4, Ksh) \text{ in}$$
$$\overline{c_{10}[i_S]}\langle RAND', SQN', MAC' \rangle;$$
$$c_{13}[i_S](RES' : prpcblocksize);$$
$$\text{if } XRES' = RES' \text{ then}$$
$$\text{let } mKNAS' = \text{mkgen}(K'_{ASME}) \text{ in}$$
$$\text{event } (endAuthS1(h, RAND', K'_{ASME}).$$
$$\overline{c_{14}[i_S]}\langle NASsma, \text{mac}(NASsma, mKNAS') \rangle.$$

**Fig. 9.** CryptoVerif formalization of serving network (eNB/MME) actions in LTE AKA+1, where $NASsma$ is considered to be a non-secret constant

$$Q_K = !^{i_K \leq N_2} c_6[i_K](Khost : sn, Kkey : key, Kmkey : mkey);$$
$$\text{let } Rkey : key =$$
$$\quad \text{if } Khost = S \text{ then } Ksh \text{ else}$$
$$\quad Kkey$$
$$\text{in}$$
$$\text{let } Rmkey : mkey =$$
$$\quad \text{if } Khost = S \text{ then } mKsh \text{ else}$$
$$\quad Kmkey.$$

**Fig. 10.** CryptoVerif formalization of the process for registering long-term keys shared between honest home network and possibly dishonest serving networks in LTE AKA and LTE AKA+1

$$Q_L = !^{i_L \leq N_2} c_7[i_L](Lhost : ue, Lkey : prpckey, Lprfkey : prfkey);$$
$$\text{let } Ukey : prpckey =$$
$$\quad \text{if } Lhost = U \text{ then } prpckuh \text{ else}$$
$$\quad Lkey$$
$$\text{in}$$
$$\text{let } Uprfkey : prfkey =$$
$$\quad \text{if } Lhost = U \text{ then } prfKuh \text{ else}$$
$$\quad Lprfkey.$$

**Fig. 11.** CryptoVerif formalization of the process for registering long-term keys shared between honest home network and possibly dishonest user equipments in LTE AKA and LTE AKA+1

$$Q_H = !^{i_S \leq N_H} c_5[i_H](h_1 : sn, e'_1 : bitstring, mac'_1 : macs);$$
$$\quad \text{find } j_1 \leq N_2 \text{ suchthat defined}(Khost[j_1], Rmkey[j_1], Rkey[j_1]) \land (Khost[j_1] = h_1) \text{ then}$$
$$\quad \text{if check}(e'_1, Rmkey[j_1], mac'_1) \text{ then}$$
$$\quad \text{let } i_\bot(\text{concat}_2(h_2, = h_1)) = \text{dec}(e'_1, Rkey[j_1]) \text{ in}$$
$$\quad \text{find } j_2 \leq N_2 \text{ suchthat defined}(Lhost[j_2], Ukey[j_2], Uprfkey[j_2]) \land (Lhost[j_2] = h_2) \text{ then}$$
$$\quad \text{new } RAND : nonce;$$
$$\quad \text{let } MAC = \text{prpcenc}(\text{concat}_8(const_1, SQN, RAND), Ukey[j_2]) \text{ in}$$
$$\quad \text{let } XRES = \text{prpcenc}(\text{concat}_9(const_2, RAND), Ukey[j_2]) \text{ in}$$
$$\quad \text{let } CK = \text{prpcenc}(\text{concat}_9(const_3, RAND), Ukey[j_2]) \text{ in}$$
$$\quad \text{let } IK = \text{prpcenc}(\text{concat}_9(const_4, RAND), Ukey[j_2]) \text{ in}$$
$$\quad \text{let } K_{ASME} = f(Uprfkey[j_2], \text{concat}_6(SQN, CK, IK, h_1)) \text{ in}$$
$$\quad \text{new } r_2 : seed;$$
$$\quad \text{let } e_2 = \text{enc}(\text{concat}_7(h_2, K_{ASME}, RAND, XRES, SQN, MAC), Rkey[j_1], r_2) \text{ in}$$
$$\quad \text{event } endAuthHU(h_2, RAND, XRES);$$
$$\quad \text{event } endAuthH(h_2, h_1, RAND, XRES);$$
$$\quad \text{event } endAuthH1(h_2, h_1, RAND, K_{ASME});$$
$$\quad \overline{c_8[i_H]}\langle e_2, \text{mac}(e_2, Rmkey[j_1])\rangle.$$

**Fig. 12.** CryptoVerif formalization of home network (HSS/AuC) actions in LTE AKA and LTE AKA+1

$$\text{main} = !^{i \leq N} c_{start}();$$
$$\quad \text{new } rKuh : keyseed;$$
$$\quad \text{let } Kuh = \text{kgen}(rKuh) \text{ in}$$
$$\quad \text{new } rmKuh : mkeyseed;$$
$$\quad \text{let } mKuh = \text{mkgen}(rmKuh) \text{ in}$$
$$\quad \text{new } r0Kuh : prfkeyseed;$$
$$\quad \text{let } prfKuh = \text{prfkgen}(r0Kuh) \text{ in}$$
$$\quad \text{new } rprpcKuh : prpckeyseed;$$
$$\quad \text{let } prpcKuh = \text{prpckgen}(rprpcKuh) \text{ in}$$
$$\quad \text{new } rKas : keyseed;$$
$$\quad \text{let } Kas = \text{kgen}(rKas) \text{ in}$$
$$\quad \text{new } rmKas : mkeyseed;$$
$$\quad \text{let } mKas = \text{mkgen}(rmKas) \text{ in}$$
$$\quad \text{new } rKsh : keyseed;$$
$$\quad \text{let } Ksh = \text{kgen}(rKsh) \text{ in}$$
$$\quad \text{new } rmKsh : mkeyseed;$$
$$\quad \text{let } mKsh = \text{mkgen}(rmKsh) \text{ in}$$
$$\quad \text{new } SQN : nonce;$$
$$\quad \overline{c_{start}}\langle\rangle$$
$$\quad (Q_U^* | Q_S^* | Q_H | Q_K | Q_L)$$

**Fig. 13.** CryptoVerif formalization of the main process, *i.e.* game $G_0$, in LTE AKA and LTE AKA+1, where, respectively, $Q_U^* = Q_U, Q_S^* = Q_S$ and $Q_U^* = Q_U^+, Q_S^* = Q_S^+$

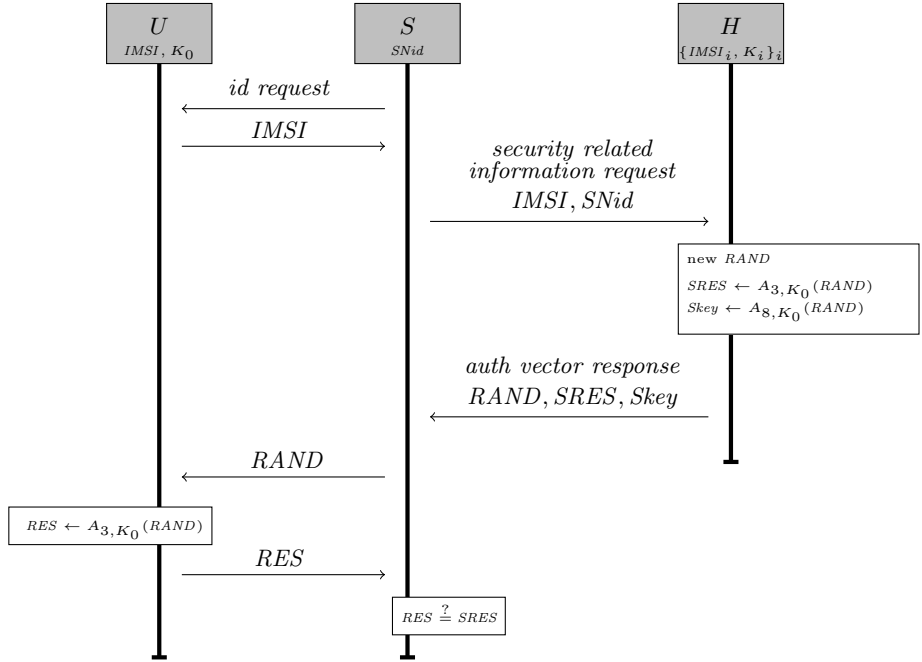

**Fig. 14.** The GSM Subscriber Identity Authentication Protocol.

SIA case, the specifications [12,11] are only concerned about adversaries that attack the radio path, *i.e.* the connection between user equipment and base stations, while completely neglecting other connections, including the connection between serving and home network. So it does not violate the specifications even if there is no protection of the authentication vector response and the session key is transmitted in the clear from the home to the serving network. An attacker that is able to listen on the connections within the core network does not need to resort to the session-mixup attack to successfully violate GSM security. We warn, however, that any GSM operators that would like to protect the connection between home and serving networks need to correct the GSM SIA protocol.